# Modeling adult skeletal stem cell response to laser-machined topographies through deep learning

**BENITA S. MACKAY,**[1,*] **MATTHEW PRAEGER,**[1] **JAMES A. GRANT-JACOB,**[1] **JANOS KANCZLER,**[2] **ROBERT W. EASON,**[1] **RICHARD O.C. OREFFO**[2] **AND BEN MILLS**[1]

[1]*Optoelectronics Research Centre, Faculty of Engineering and Physical Sciences, University of Southampton, Southampton, United Kingdom, SO17 1BJ*
[2]*Bone and Joint Research Group, Centre for Human Development, Stem Cells and Regeneration, Institute of Developmental Sciences, Faculty of Medicine, University of Southampton, Southampton, United Kingdom, SO16 6HW*
*\*b.mackay@soton.ac.uk*

**Abstract:** The response of adult human bone marrow stromal stem cells to surface topographies generated through femtosecond laser machining can be predicted by a deep neural network. The network is capable of predicting cell response to a statistically significant level, including positioning predictions with a probability $P < 0.001$, and therefore can be used as a model to determine the minimum line separation required for cell alignment, with implications for tissue structure development and tissue engineering. The application of a deep neural network, as a model, reduces the amount of experimental cell culture required to develop an enhanced understanding of cell behavior to topographical cues and, critically, provides rapid prediction of the effects of novel surface structures on tissue fabrication and cell signaling.

## 1. Introduction

With an approximate annual cost of £2.1 billion for osteoporosis to the UK National Health Service (NHS) and costing between 1.0% and 2.5% of gross domestic product for westernized countries, osteoporosis and osteoarthritis (OA) represent major socio-economic challenges in an aging demographic [1, 2]. Indeed, OA is the most common form of arthritis worldwide and to date there is no definitive cure for this debilitating disease [3]. Current approaches to alleviate this skeletal disease include pain medication, bone grafts/stem cells and implants. The former solution is unsustainable – a recent public health report showed 5% of UK citizens are prescribed opioids [4], which have limited long-term benefit and severe issues including lack of clinical proof of pain reduction and numerous associated risks of opioid use [5]. The use of grafts/stem cells and implants are not without risk and include the possibility of rejection. Hence, there is a growing need for innovative techniques to promote implant integration and reduce the failure rate of osteopathic intervention.

The cell type responsible for bone formation, the osteoblast, is derived from a multipotential marrow stromal stem cell. In order to create innovative techniques for skeletal repair, a vital first step is to advance the understanding of adult bone marrow-derived stromal stem cell development; this is achieved here through the use of femtosecond laser machined topographies and deep learning. Harnessing topographical cues offers an accepted and promising technique to control stem cell fate and function, as cells respond to the shape of their environment due to changes in contact guidance, cell spreading and contact inhibition [6]. Cell behavior can therefore be influenced through the topographical engineering of surfaces and the use of surface-directed biotechnologies [7, 8]. Variations in surface topography have been demonstrated in a raft of studies to exert a number of physiological effects including; i) cell adhesion [9, 10], ii) density and spreading [11, 12], iii) cytokine secretion (important to cell signaling) [13], iv) proliferation and v) skeletal stem cell differentiation [14]. Cell behavior can therefore be influenced through the precise engineering of surfaces and use of surface-directed biotechnologies, as shown in Fig. 1, where cell positioning is dependent on surface topography. Through the generation of laser-machined microtopographies (structures on ~10 μm scale) and assessment of the cell response to laser-ablated glass, deep learning can predict stem cell behavior and provides a model and platform for further stem cell behavioral investigation in the absence of extensive cell culture management and analysis.

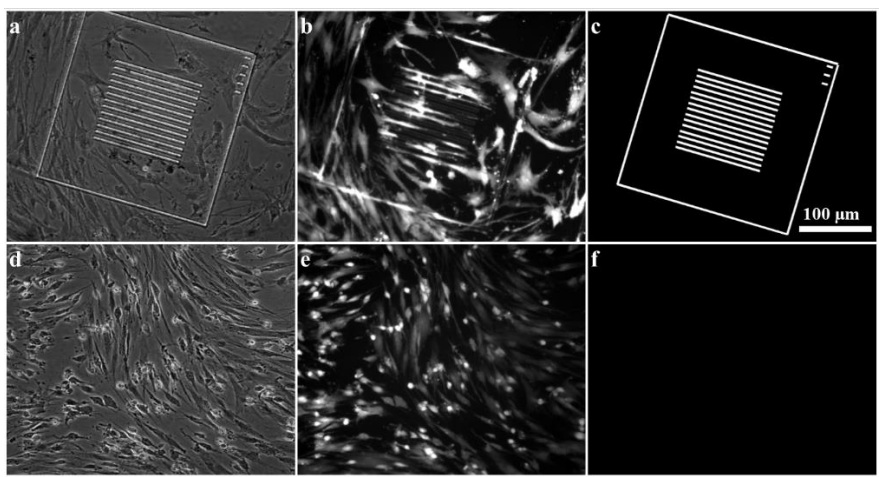

Fig. 1. Adult skeletal stem cell alignment and adhesion parallel to microscale laser machined lines taken with brightfield (a) and fluorescent (b) microscopy in response to (c) the surface topography, seen as an array depicting laser machined areas (white) on an otherwise smooth topography (black). This is compared to (d, e) adult skeletal stem cell positioning on (f) the smooth surface without altered topography. Scale bar in (c) applies to (a-f).

Photolithography has been used with significant success in the creation of topographies capable of inducing directed cell response, including adult skeletal stem cell differentiation [14, 15]. However, mask designs must be carefully planned in advance and the process can be time consuming and costly, especially if many iterations of the design are necessary. Whilst photolithography is most economical and suitable when many identical copies of the same pattern are required, femtosecond laser machining can be more appropriate when it is necessary to rapidly assess the performance of a small number of new surface topographies. This is the mode of operation that is eventually envisaged for the deep neural network model. The neural network model (for prediction of cell behavior) may be used to select the most promising candidate surface topographies; both the neural network model and the surface topography design can then be iteratively improved with feedback from a small number of experimental measurements. In this case, a priori fabrication of all possible masks is avoided, and the number of experimental measurements needed is greatly reduced. Laser machining has been applied in previous works to form single and low parameter space topographies, such as spikes and pits, to determine cell response [16, 17]. The current study sets out to examine the ability of adult bone stem and progenitor cells to detect discrete alterations in surface topographies, and which combination of factors such as cell adhesion, cell morphology, intercellular contacts and cell cycle progression, could be exploited for cell control. Given the fact that complex signaling sequences that instruct cell behavior depend on these various parameters, it is likely that simultaneous alterations to several parameters will be required to modulate cell function. The ability to generate and predict such a complex, large, parameter space, to modulate and control cell function, is currently beyond human and simple computational ability.

Deep learning has been used in recent years to achieve beyond human solutions and enhancements in the healthcare field, for example as evidenced with super automated focusing in microscopy data and platelet detection in diluted whole blood samples [18-20]. Deep learning has also been applied with repeated success to the biomedical imaging field [21-25]. Recent advances in deep learning show that, when given enough sufficiently varied training data, the need for a complete sampling of parameter space can be unnecessary [26-28]. Thus, as shown here, training a deep neural network on varied, yet limited, topographies and the subsequent cell response can result in predictions of cell response on topographies unseen by the network and untested in a laboratory setting. This approach could then lead to the determination of optimal topographies without the need for experimental analysis and the time and cost implications therein. Critically, such a platform offers new approaches to derive insights into stem cell behavior. A deep neural network, applied as a model to approximate cell response, facilitates the modulation of multiple parameters and cell response analysis enhancing scientific understanding. This model can then be used to create results that follow rules derived from experimental data at pace, on a larger scale far exceeding experimental strategies and thereby provide enhanced understanding and development of cell behavior mechanisms.

## 2. Method

The methodology utilized is divided into two sections, with details of the experimental set-up for generation of training data in section 2.1, including the laser machining of topographies, cell growth and subsequent imaging. Details for the establishment of the deep neural network are presented in section, 2.2, including information on the network architecture and relevant hyperparameters.

*2.1 Experimental set-up*

To generate a deep neural network that is capable of predicting cell response when inputted with a laser-machined topographical pattern, the relevant training data must be produced (shown in Fig 2). The input topography to the network (image 1), was laser machined onto a glass sample (image 2) prior to cell culture on the laser-machined topography. Images were captured at varying time points (image 3), to facilitate matching of the fluorescent images of cell position on a sample to the corresponding topography pattern (image 4). Images 1 and 4 provided the input and output pair required by the deep neural network. The deep learning platform would subsequently transform the input, a laser-machined topography image (image 1), to the output of a corresponding fluorescent image depicting the cell position (image 4), without the need for either laser machining or cell culture.

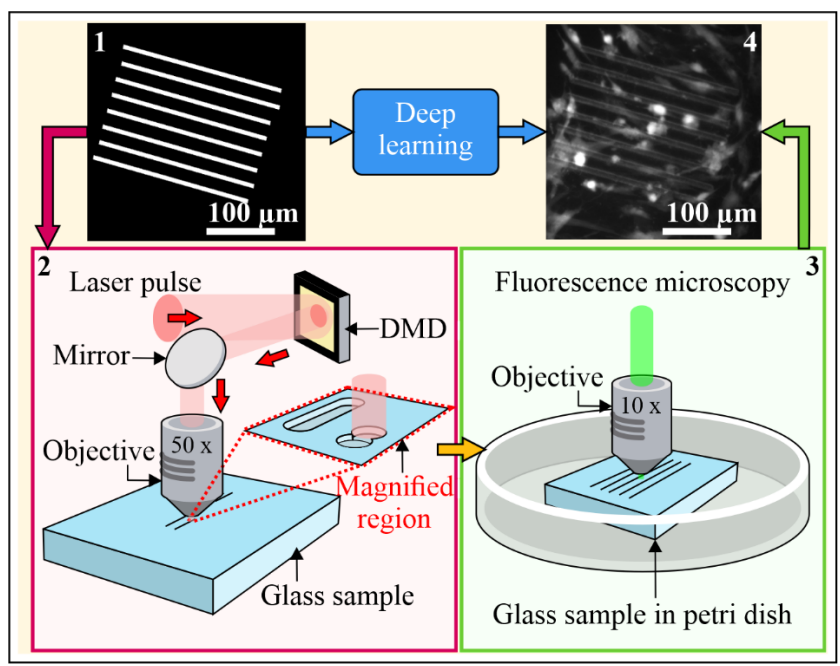

Fig. 2. Method for generating a model for cell response. Step 1 topography design; step 2 laser machining of the topography onto a glass sample; step 3 image cells grown on the adopted topography to determine cell response and step 4, process and align the images to the corresponding input topography in step 1. Deep learning can be used to predict image 4 from image 1, without the need for steps 2 and 3.

Glass coverslips served as a convenient substrate for this experiment primarily because of the known ability of cells to adhere to glass. Femtosecond laser machining of glass, taking advantage of nonlinear absorption, is also known to allow fabrication of features with high fidelity. Soda-lime glass coverslips were used in these studies as these coverslips are less susceptible to fracture than pure silica glass post-machining.

For laser machining, a 1 W Ti:sapphire, 150 fs pulse duration laser, with a 1 kHz repetition rate, centered at 800 nm, was used. The laser pulses were spatially shaped using a digital micromirror device (DMD), to control the spatial intensity of the laser pulses on the sample, and therefore the topography of the ablated structures [29]. While any shape could in practice have been used, circular patterns on the DMD were chosen to produce circularly shaped laser pulses on the surface of the substrate. When combined with substrate movement (via a 3-axis translation stage), the result was a continuous ablated line, with line thickness corresponding to the diameter of the projected circle shape. The size of the circle pattern on the DMD was optimized in order to produce a specific ablated line width. In this case, a DMD was used as it offered rapid digital switching between different line widths for topographical variation. Made up of a 604×684 array of square ~7.6µm mirrors (DMD pixels), the DMD is utilized to create specific spatial intensity profiles for sample ablation. The use of a Pi-shaper permitted a uniform intensity profile necessary for consistent topographical patterning of the sample, and, alongside the DMD, will allow for consistent machining over curved samples, such as bone samples and titanium implants for future expansion and increased versatility of the network model [30].

With a circle pattern set on the DMD of radius between 75 and 100 pixels, and the laser power reduced to 400 mW (controlled using a variable density filter) before input to the DMD, laser machined lines resulted in a typical thickness range of 7.5 to 12 µm on the glass sample. Square boxes were machined with a dimension of 500 µm on a side to create a boundary between individual topographical patterns at a size where, during imaging, there was one box per image at a 10x magnification. This ensured that both the alignment of cells in one box was less likely to be influenced by the cell alignment in an adjacent box and also aided image capture and alignment. Three different laser-machined patterns (parallel lines at different separations within the machined square boxes) have similar microscale patterning, but varied nanoscale patterning, shown in Fig. 3.

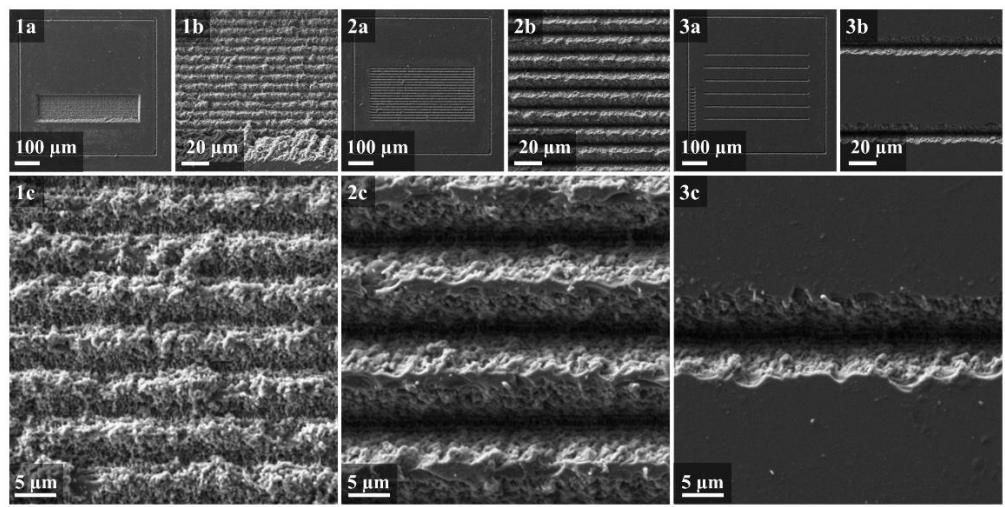

Fig. 3. Three different laser-machined patterns on a glass coverslip, imaged with a scanning electron microscope (SEM). 1a-c has lines machined with a separation of 5 µm: close enough together that the deeper ablated lines, seen as darker grey and black in 2a-c and 3a-c, are not produced. Instead, disordered nanoscale features are present in uniform microscale lines. A larger separation of 10 µm (2a-c) or greater (3a-c) creates larger variation in depth, with clear ablated lines surrounded by disordered nanoscale variation. Areas not laser machined are smooth, with virtually no microscale features.

Following generation of the chosen topographies and sterilization of the coverslips in ethanol, cells were cultured on the substrates at 37ºC and 5% $CO_2$ atmosphere in medium (Alpha Minimum Essential Medium Eagle supplemented with 10% fetal calf serum and 1% penicillin/streptomycin), with media changed every three days. Cell cultures were imaged at set time points from day 0 until cells covered the coverslips (fully confluent), to a maximum of day 30, and dependent on cell seeding. Multipotent skeletal progenitor enriched populations from human bone marrow, isolated using a Stro-1 positive antibody, were used in these studies. In brief, human bone marrow aspirates were collected from hematologically normal patients undergoing routine elective hip replacement surgery. Only tissue samples that would have been discarded were used following informed consent from the patients in accordance with approval from North West - Greater Manchester East Research Ethics Committee (Ref-18/NW/0231). Multipotent skeletal stem and progenitor cells were enriched from bone marrow aspirates following our standard protocols [31, 32].

The cells were fluorescence imaged with Vybrant cell tracker dye via a Nikon Eclipse microscope at different timepoints post seeding. Variation in cell seeding density provided a useful randomization such that the network was not constricted to predicting a specific cell density, and cell response could be determined for a range of densities. While all images were taken with an equal exposure time, any alterations in brightness were compensated for through processing to adjust for contrast and brightness prior to use as training data for the network. This ensured that values for cell density and for cell positions, which were derived using image recognition techniques, were consistent between measurements. In total, 203 fluorescent images from six donors were used to train the network.

Testing data were processed using the same procedures as for training data. However, these images were not used in training and remained unseen by the network until the network was fully trained. Critically, rather than randomly extracting a percentage of data from the training data and applying this for testing, a completely new dataset was used, with cells derived from a different patient. The new data were used as input to the trained network and the output subsequently analyzed in comparison to the real experimental images to determine the validity of the network as a model for Stro1+ enriched skeletal stem cell response to laser-machined topographies.

*2.2 Set-up of the deep neural network*

The deep neural network architecture consisted of both a generator and a discriminator network [33], where the former follows a W-Net architecture [34] (Fig. 4), and the latter was a convolutional network without deconvolution. The W-Net architecture is based on the U-Net architecture, increasingly common in biomedical fields [35-37], with a secondary U-Net included in the architecture for improved performance, based on deep cascade learning [38]. The input to the W-Net consisted of three data channels: a topography channel corresponding to the laser-machined topography (A); a time channel corresponding to the time point at which the output image would have been taken (day 0, 1, 8 or 30) (B); and a cell density channel (C), which was a randomized array corresponding to the brightness of the original, unprocessed fluorescent image and therefore only an approximation to the cell density. The input channels were individually altered at every iteration, together with the corresponding target output, which depended on the unique topographical pattern, cell density and time point at which the target output image was obtained. Examples of patterns used in training the network are shown in Fig. 3.

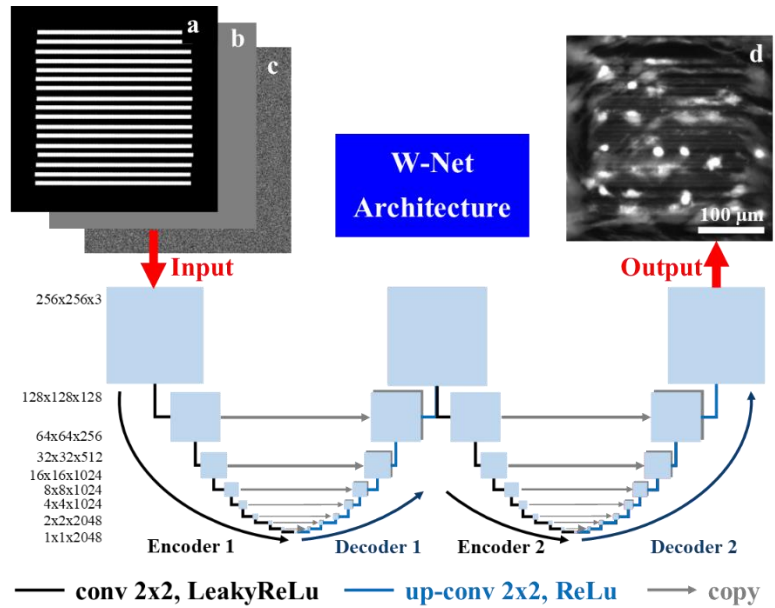

Fig. 4. The deep neural network W-Net architecture consists of multiple convolutional layers and skip connections between encoder and decoder sections. The input contains three data channels: topography (a), time (b) and density (c). The output (d) shows the neural network prediction of cell growth (as it would appear under a fluorescence microscope) for the topography, timepoint and randomized cell density seed shown in A, B and C respectively.

The network operated at a resolution of 256 x 256 pixels to minimize both data loss and time spent training, where images were reduced in size through randomized cropping, from a maximum 1280 x 1024 to 512 x 512, before being resized to 256 x 256 by the network. The network was trained for 25 epochs (where one epoch is defined as training on all training images exactly once) with a learning rate of 0.0005 and a batch size of 1, which took two weeks on a NVIDIA Quadro P6000 GPU. The generator network was based on an encoder-decoder architecture, with 34 layers, a stride of 2, a 4 x 4 kernel size and used rectified linear activation functions. This resulted in image size decreasing from 256 x 256 down to 1 x 1, then increasing back up to 256 x 256, then repeatedly decreased and increased. The generator also contained skip connections between the mirrored layers, illustrated by the grey arrows in Fig 3. The discriminator was formed of 4 layers of convolutional processes with a stride of 2, taking the image size from 256 x 256 down to 32 x 32, leading to a single output, via a sigmoid activation function that labelled images as realistic or unrealistic. After a training iteration (where a single image is input through the network), outputs from the network were compared to real labelled images, leading to network improvements achieved via backpropagation. At each iteration, the discriminator network was inputted either the generated image or the manually labelled image, which the discriminator would correctly or incorrectly identify. The higher the discriminator error in identifying the authenticity of the image, the better the generated image, so this was used to help train the generator. By appropriately weighting and combining both the backpropagation and discriminator output, which was altered throughout training to encourage realistic images and to statistically correct images at different stages, the generator was trained to produce realistic fluorescent images of cells in the most statistically likely position, rather than a blurred image that combines all likely cell positions.

## 3. Results and Analysis

Following the establishment of the network using a defined training data set, test data were added into the network, as previously undertaken with training data (Fig 3). The outputs from the network were subsequently compared to experimental images for the corresponding topography, cell density and time point. Cell density refers to the desired density of cells in the image generated by the network and is not related to initial seeding densities. After the output had been analyzed and observed to be statistically significant, validating the network, the network was used as a model to determine the minimum line separation required for cell alignment. This result was subsequently compared to experimental data to determine the success of the deep neural network to determine and model the skeletal stem cell response.

### 3.1 Capability of the deep neural network

To determine the input parameter range that produced a visually realistic output by the network, a simple test of cell density and time-point for a blank (unmachined), topography was conducted (Fig. 5). The input parameters were altered individually, to ensure the network was modifying predictions independently to each input channel.

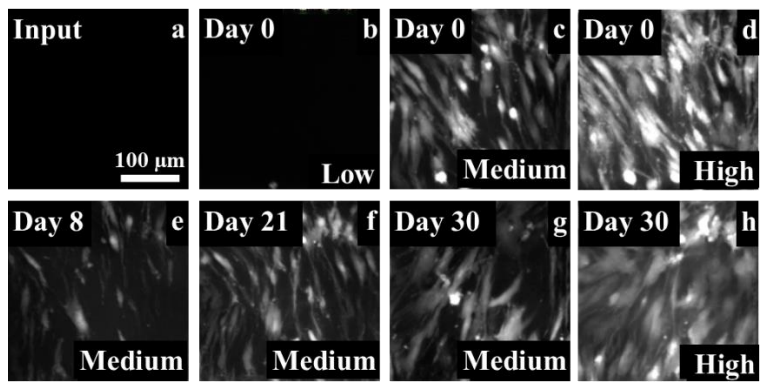

Fig. 5. Testing the independence of time and density channels, an input without laser-machined topography (a) is used to compare network output while other parameters vary, including

variation of density for a set time point (b-d and g-h) and variation of time points for a set density (c, e-g) and (d, h). Low, Medium and High labelling at the bottom of images (b-h) indicate the cell density input into the network, relative to confluency, while the time points are labelled at the top left. Images (b-d) show how increasing density while time is static results in different outputs. Images (c, e-g) show how increasing time while density remains unchanged results in different outputs, both from each other and from (b-d). Image (h) shows the result of both a high time point and density. Scale bar in (a) applies for (a-h).

When the cell density and time input channels were independently altered, the network was able to make a variety of output image predictions, for the same blank topographical input. Fig. 5 (a-d) demonstrates how variation to the input density channel affects the network output prediction. At the time-point of day 0, Fig. 5 (c-d), the skeletal cells are clearly distinguishable, with numerous spherical, brighter cell clusters and structures present. As the time-point increased (Fig. 5 e-g), fewer bright and spherical cells could be observed, as the skeletal populations adhered to the surface and spread, displaying typical behavior that was learned from the experimental training data. Differences in cell number across Fig. 5 (c, e-g) are a result of the fluctuations in individual pixel value for density input, even while the mean pixel value for the density input channel remains uniform. While an increase in density and an increase in time-point could both be incorrectly oversimplified to an increase in cell number, Fig. 5 (d) and (h) show the network successfully created images that vary across the independent input parameters, as the generated output images were different.

Furthermore, to confirm that changes to the density input channel result in a realistic and varied density in the network predictions, a test of two simple input topographies, parallel lines and crossed parallel lines, was undertaken (Fig. 6). For each increase in input density, a visible increase in the cell density was evident. In contrast to changes in time points in Fig. 5, where there was no visible repetition of a cell position, cell position in lower density images in Fig. 5 and Fig. 6 were typically repeated in higher density images. Such repetition in cell positioning was the impact of both input density and topography, with the network modifying the output to control not only for cell density but also for probabilistic cell position on a given topographical pattern. A lower density input provided the most statistically likely cell position without flooding the image with additional cell positions, an issue that could impact the effectiveness of the statistical analysis. This lower density approach was exploited in the next section, in which the statistical validation of the network was evaluated.

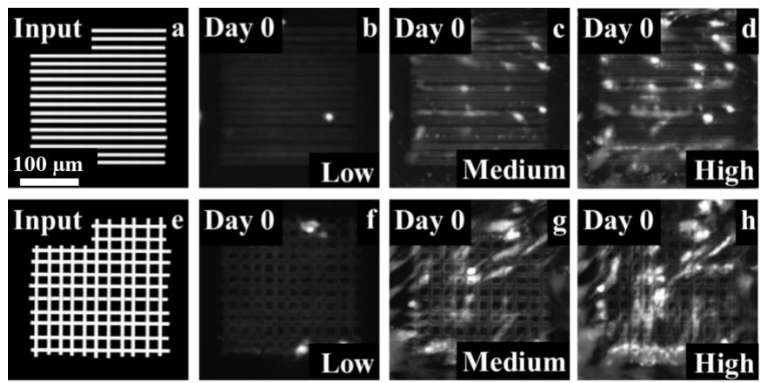

Fig. 6. Testing the connection between input topography and density channels: two topographical inputs are used, one of parallel lines (a) and another of crossed parallel lines at right angles (e), to compare network output while the time input channel remains unchanged (b-d, f-h). Low, Medium and High labelling at the bottom of images (b-h) indicate the cell density input into the network, relative to confluency, while the unchanged timepoint of Day 0 is labelled at the top left. Scale bar in (a) applies for (a-h).

As the parameter space was larger for the topographical channel than the time and density channel, parameter space was initially tested independently of both time and density input (Fig.

7). For a set time point of day 0 and a medium input density, with the cells not reaching confluency but covering the image space, the separation between parallel lines was increased from 20 µm to 290 µm at randomly generated intervals, and predictions of cell behavior were altered as a result. At 20 µm separation, areas of cell alignment and areas where cells appeared to stretch across the lines were observed. However, at 90 µm separation, most cells were observed adhered onto the lines, with fewer cells branching out across multiple lines. Interestingly, some cells not adhered to the machined lines were noted to be parallel. The exact mechanisms behind this cell alignment, such as cytoskeletal organization and interaction with other cells, has been observed as a response to various topographical cues [39-45]. At a wound interface, cells have a migratory phenotype switched on rather than a proliferative (division) phenotype and these cells influence the adjacent cells, sending signals to migrate to the wound space. Cells adhered to the machined lines could be releasing signals to the unconditioned cells to align parallel to the machined lines. This behavior is important in development of angiogenesis, biological neural networks and the growth plate (chondrocyte columns) in the growth of long bones. Cell behavior remains relatively unchanged for 120 µm but at 290 µm separation, while the skeletal cells appear to demonstrate a preference to adhere to the machined lines than to unmachined areas, there was relatively little cell alignment. This network predicted preference for machined areas, supported by experimental data (Fig. 1), implies that the microscale texturing of the laser machined glass promotes adhesion and/or reduces cell mobility.

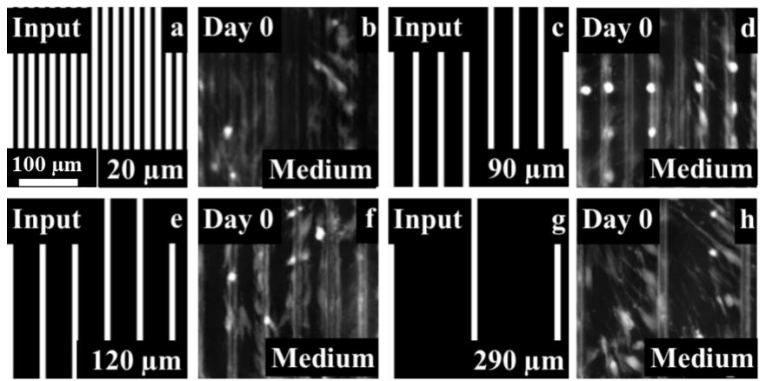

Fig. 7. Testing the connection of input topography only, the time and density channels remaining unvaried. The input topographies are parallel lines of uniform 25 µm width at varying separation (a, c, e, g) and the output is the predicted cell positioning (b, d, f, h). Scale bar in (a) applies for (a-h).

Critically, a significant change in cell behavior was observed when only the separation between lines in the input topography channel was varied: Differing input topographies result in changes to elongation, position and interaction. Together, Fig. 5-7 demonstrated that the output of the neural network changed when the input channels of time, density and topography were independently altered. These results demonstrated that the network could produce varied output images for a range of input parameters. However, a larger input range beyond topographical patterning consisting only of straight lines with thickness in the range 10 µm to 25 µm was required for full network functionality. Therefore, the functionality of the network was also tested using circles, curves, and patterns with a greater range of line width (Fig.8). Although the network had only previously been trained on a line width of approximately 10 µm, the lines overlapped in such a manner to introduce thicker lines into the training data. Importantly, lines substantially thinner are new to the network.

The current results showed that the network could predict cell responses for these new topographical inputs, including cell interaction with concentric circles, 1b, and cell alignment with curved lines, 2b, despite no data input from curves in the training data (Fig. 8).

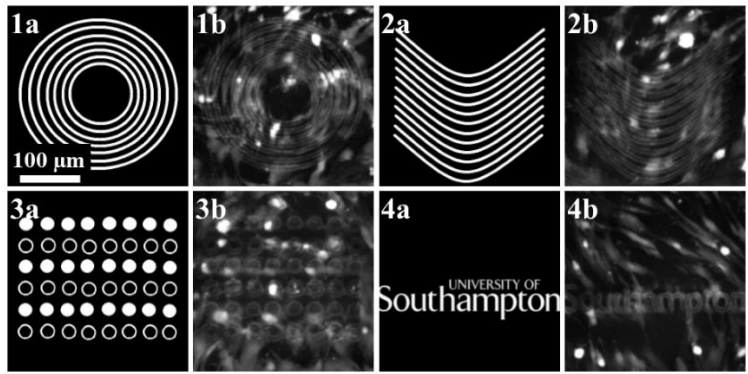

Fig.8. Extending the range of topographical parameters for input to the network. The network was examined with concentric circles in pair 1, curves in pair 2, filled circles in pair 3 and alphanumeric characters in pair 4. Each pair consists of the topographical input (a) and the network predicted output (b). Scale bar in (a) applies for (a-h).

The topographical input, 3a, and network predicted output, 3b, of filled circles and open circles is significantly different from patterns encountered by the network during training. Nevertheless, the current results showed the network could predict that there was higher cell interaction with the filled circles over thin open circles, showing the preference for cells to adhere to the laser-machined surface in defined topographical patterns (Fig. 8), as seen previously in both network prediction (Fig. 7) and experimental data (Fig. 1). As expected, a range of patterns, without parallel structures, do not result in cell alignment or consistent adhesion, which the network was able to successfully predict, evidenced by the data from the topographical input of alphanumeric characters, and from the cell (non)response to writing in pair 4a/4b – some of the machined areas are barely visible where there is no cell activity.

### 3.2 Validation of the deep neural network

The testing dataset used to validate the network contained unseen images of cells (cells from a distinct patient, on unseen topographies, at an unseen cell density, and contained unseen time-points). The output of the network, a prediction of cell response to a given input topography, density and time-point, was compared to the experimentally obtained images for the same input parameters (Fig. 9 a-d). Fig. 9 (a) shows the topographical input to the network, (b) shows the network prediction for cell position and (c) shows the real experimental positioning of cells on the same topography. (d) is a comparison image, where blue pixels indicate the network prediction, red pixels indicate the experimental position and green pixels show the areas were both prediction and experimental positions coincide. Fig. 9 (e-h) show the comparison images with topographical input (a) superimposed as translucent white lines for easier visual analysis. Fig. 9 (e-h) indicate areas of green pixels, where the network correctly predicted there would be skeletal cells present. (e) shows cells correctly predicted to be present in the area of the laser-machined lines and aligning midpoint along one line at day 0 for low density. (f) and (g) show testing comparison images for days 8 and 15 respectively, where cell activity was correctly predicted along the border lines, and (h) shows the ability of the network to correctly predict the likelihood of a single cell being positioned on a topographical line. Although the network prediction of the cell density was higher than for all experimental images, evidenced by the increased level of blue to red, this variance was a result of a relatively low cell density within the testing images, even at later time-points, which was novel to the network.

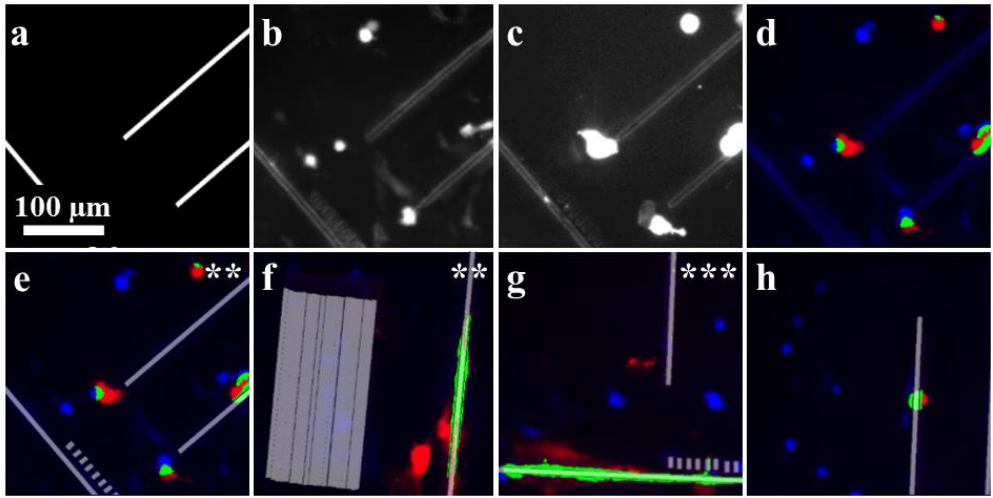

Fig. 9. Successful testing of the neural network using cells acquired from an unseen patient adhered to an unseen topography for validation. A randomly selected input topography (a) is input into the neural network and the predicted cell positioning (b) is output by the network. This is then compared to (c) to produce a comparison figure (d), where blue is the output network, red is a real cell positioning and green is areas of agreement. E-H are comparison images of cell positioning with transparent grey lines showing the laser-machined areas and input topography. (e-g) are statistically significant, where (e-f) $P < 0.01$, shown with **, and (g) $P < 0.001$, shown with ***. Scale bar in (a) applies for (a-h)

To determine whether these results were significant, each image was statistically analyzed using the hypergeometric distribution probability mass function [46] to evaluate the likelihood that such an answer could be achieved through random positioning of cells across the image. Therefore, each image was thus divided into 256 approximate cell-sized sections, rather than using exact pixel to pixel comparison. The probability was then calculated for the likelihood to correctly position cells (green in Fig. 9) in the corresponding sections filled with cells (red in Fig. 9) in the experimental image. The total number of cell positions (blue and red in Fig. 8) were determined computationally to reduce bias. Fig.9 (e-g) were observed to be statistically significant, with (e-f) $P < 0.01$ and (g) $P < 0.001$. Paradoxically, while predicting the exact position of a cell on a single line may initially appear to be an unlikely result, (h) was not statistically significant ($P = 0.055$) due to increased network predicted density and consequent binomial probability adjustments. However, (h) showed that a network bias towards higher density did not limit the network from predicting cell positioning. The correct prediction of a singular cell positioning along the single machined line in (h), alongside statistically significant predictions in (e-g), showed that the positioning of a cell is not inherently random, even when adhering to a single straight line.

The current findings indicate that, with a greater emphasis on levels of cell density examined, the network may be able to produce highly statistically significant results for all input parameters, since higher density network outputs for both cell positioning and behavior were shown in Fig. 5-7. Furthermore, when the probability of placing individual random pixels, instead of cell-sized areas, was calculated, all outputs (e-h) were statistically significant, with $P < 0.001$.

Negative controls, no cells or no topography, were introduced to eliminate the possibility of overfitting, the network effectively "memorizing" the training data (Fig. 10). Unlike for the previous testing, a difference between the predicted and experimental images was viewed as a success. Pairs 1 and 2 show the network prediction (1a/2a) and experimentally obtained (1b/2b) images for a cell density input of zero. The approximated cell density was observed to be virtually null, yet there were discrete differences in both pairs, as 1a showed a single cell and

1b displayed no cells. In 2a, a higher level of fluorescence in the laser machined lines on the glass substrate than in 2b was observed. This deviation was noted to be a consequence of the buildup of fluorescent particles in the laser-machined lines, giving a cell density approximation input just above zero. Currently, the network could not determine whether the low fluorescence in extremely sparse images with one cell or less was due to low cell density or a buildup of fluorescence in the laser machined lines, and therefore randomly assigned an output of either low cell density or fluorescent debris in lines. This limitation in low fluorescent and single cell imaging is an area for progression in future work.

Pairs 3a/3b and 4a/4b in Fig. 10 illustrate the differences in predicted and experimental results for an input without a laser machined topography. Pair 3 shows that, for an early time point and lower density approximation, cell shape and number were appropriately approximated. However, the positions were random in the absence of topographical cues to generate cell positioning. Pair 4 shows that, for a longer time point and higher fluorescence giving a higher cell density approximation, there were too many parameters for the network to predict a matching prediction to experimental result. These differences highlighted that results of the "negative controls" study confirmed the network did not overfit to the training data.

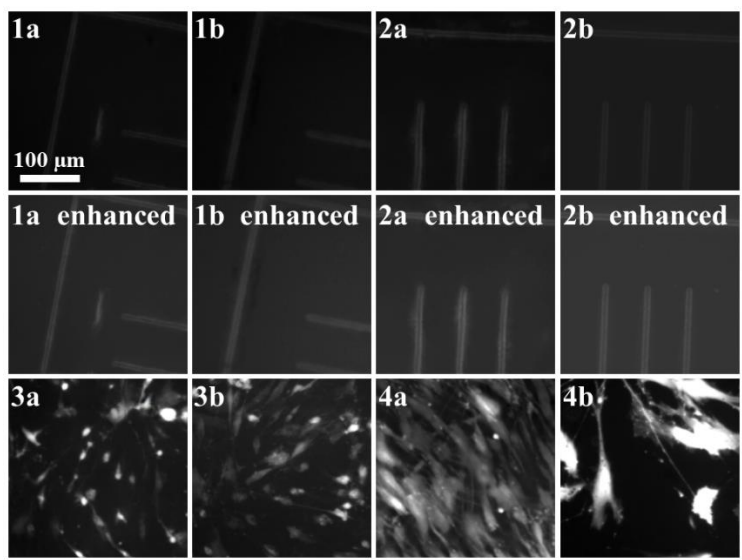

Fig. 10. A series of network predicted (a) and experimentally imaged (b) pairs. The top row, pairs 1 and 2, are for a cell density of virtually zero and the bottom row, pairs 3 and 4, are for an area with no topographical patterning. The central row is a copy of pairs 1 and 2 with enhanced contrast and brightness for visual clarity. Scale bar in 1a applies to all images in this figure.

The current data, in combination with the results of section 3.1, confirmed the successful validation of the network and indicated a potential for the network to predict cell responses for parameters not included in the training data. Thus, the network can be used as a model for cell response to new and untested topographies.

### 3.3 Using the deep neural network as a model

As the network had been found to produce statistically significant predictions of cell positioning for a given topography, the network was used as a statistical probability tool to model cell responses to unmachined, unseen, topographies. Alongside this statistical success, it has generated images that appeared to match biological observation. A possible implementation for the model is to derive alignment limitations, as shown in Fig. 11. A set of increasingly thick and separated lines were input into the network for a range of time-points and cell densities. The aim was to determine the minimum line separation for different line thicknesses, with the minimum separation required for cell alignment averaged for each line thickness and plotted on a graph.

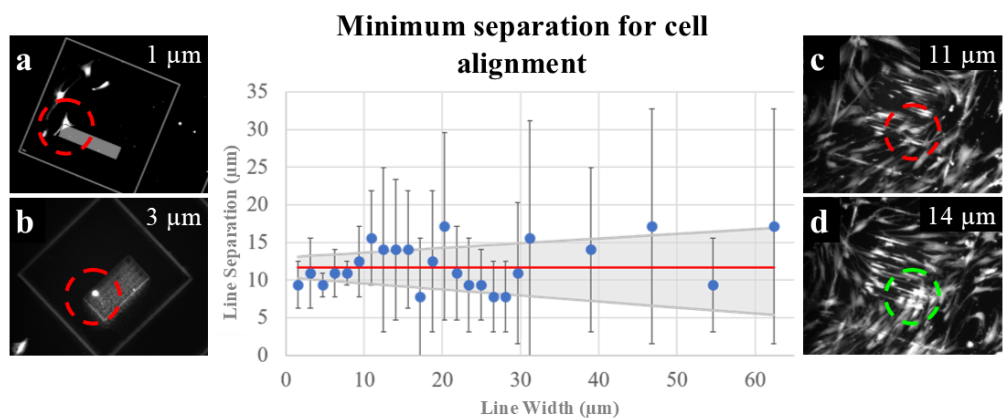

Fig. 11. A graph of minimum separation for cell alignment, obtained using model predictions from network generated images. The red linear trend line shows the (lack of) relationship of line separation in respect to line width, blue dots are the average minimum line separation for a given line width resulting in cell alignment and the grey shaded area is the error in the red trend line. The surrounding figures are experimentally obtained fluorescent images for varying line separation where there is and is not cell alignment.

The blue dots on the graph in Fig. 11 are the average minimum line separation required for cell alignment (error bars as grey lines). These error bars represent the discrete differences in minimum alignment for different time points and densities but also the difficulty in calculating minimum cell alignment. Using only network predicted images, the minimum line separation for cell alignment was independent of line width, shown by the solid red linear trend line, with a gradient of 0.0 ± 0.1 [µm/µm]. The grey shaded area shows the error for this prediction. To validate this model prediction, a value extracted from the data for minimum separation was obtained, 11.7 ± 1.3 µm, and compared to experimental data of cell alignment on machined lines. Fig. 11 (c), containing parallel lines of 11 µm line separation, showed an experimental image where cell alignment occurred in some parts of the parallel lines, but not in the area circled in a red dashed line. Fig. 11 (a) and (b), with the lines separated by 1 µm and 3 µm, both below the predicted minimum value, showed no alignment whereas Fig. 11 (d), with a separation of 14 µm, had clear cell alignment at all points along the parallel lines in the center of the image. This alignment included the area in the green dashed circle, which was not aligned at 11 µm (in the dashed red circle). This minimum observed separation was likely to be cell specific and therefore surface topographies with parallel laser machined lines can be used to visually determine different cell morphologies within the same colony without the need for invasive dyes and staining, as different cells respond to different topographical cues [15].

It is important to note, there will likely be realignment when nanotopographical surface cues interact with different cell signaling mechanisms [47], which do not play a role in these machined cell lines, given the rough texture created from laser ablation that likely promotes cell adhesion over cell alignment. Through combining parameters for nanoscale and microscale alignment cues, the behavior of cells can be fully investigated and the dominance of particular cell signaling method(s) understood. Thus, by expanding the training data to include varied materials, scales of topographical patterning, and a wider range of cell morphologies, this method of modelling cell behavior could be used to expand the knowledge of skeletal cell response to topographies to produce a universal skeletal stem cell predictor model. Further work should also focus on a larger range of donor sources to mitigate impact of natural variations between donor cells, such as proliferation variation and percentage of non-responders to osteogenic differentiation, and to increase the predictive capability of the model.

## 4. Conclusion

Utilizing topographical cues offers a promising technique to control stem cell fate and function, as cells respond to the shape of their environment due to multiple signaling pathways. Cell behavior can therefore be influenced through the topographical engineering of surfaces, which is a field that still requires extensive investigation due to the complex parameter space and intricacy of cell responses. To fully investigate a wider portion of this parameter space than is currently possible through conventional computational and manual experimental processes, a novel solution was required.

The application of a deep neural network, trained on a discrete but varied dataset of 203 fluorescent images, generated a model capable of predicting cell response to a statistically significant level. It was able to generate outputs that varied when inputs of timepoint, cell density and surface topography were both dependently and independently altered, without evidence of overfitting, as inputs unseen during training still resulted in realistic generated images. Additionally, the model had the potential to derive the minimum line separation required for skeletal stem cell alignment, validated by experimental data, of $11.7 \pm 1.3$ µm.

A deep neural network, as a model, reduces the amount of experimental cell culture required to develop an enhanced understanding of cell behavior to topographical cues and, critically, provides new avenues to explore, interrogate and define structures to modulate cell signaling with implications for a regenerative framework.

## Funding

BM was supported by an EPSRC Early Career Fellowship (EP/N03368X/1)

Research funding to RO by the Biotechnology and Biological Sciences Research Council (BB/P017711/1) and the UK Regenerative Medicine Platform Acellular / Smart Materials – 3D Architecture (MR/R015651/1) is gratefully acknowledged.

The RDM data for this paper can be found at DOI: 10.5258/SOTON/D1395

## Acknowledgements

The authors would like to acknowledge Julia Wells and Kate White within the Bone and Joint Research Group, for their constant patience and provision of technical expertise. Additional appreciation to EPSRC for funding and the Faculties of Engineering & Physical Sciences and Medicine at the University of Southampton for permission to use all required equipment and materials to complete this project.

We gratefully acknowledge the support of NVIDIA Corporation with the donation of the Quadro P600 GPU used for this research, donated through NVIDIA GPU Grant Program.

## Disclosures

The authors declare no conflicts of interest.